\documentstyle[psfig]{mn}

\newcommand{\itl}[1] {{\it #1}}

\input epsf
\def\DIRFIGS{figures/}

\newcommand{\cidfig}[6]{
    \protect\centerline{
    \epsfysize=#1\epsffile[#2 #3 #4 #5]{#6}
    }}

\newcommand{\mboxs}[1]{\mbox{{\scriptsize #1}}}

\title[Monthly Notices: Radiative shocks in galaxy formation (II)]
  {Radiative shocks in galaxy formation.\\II: Effects of a metagalactic
  UV background.}
\author[M.I. Forcada-Mir\'{o}]
  {M.I.~Forcada-Mir\'{o}$^{1,2}$\\
  $^1$Institute of Astronomy, Madingley Road, Cambridge CB3 0HZ,
  England\\
  $^2$Max-Planck-Institut f\"{u}r Astrophysik, 
  Karl-Schwarzschild-Stra\mbox{\ss}e 1, 85748 Garching b.\ M\"{u}nchen, 
  Germany
  }
\date{Accepted ????. Received ????}
\pagerange{\pageref{firstpage}--\pageref{lastpage}}

\begin{document}

\label{firstpage}

\maketitle

\begin{abstract}
I use a 1-D Lagrangian code which follows both a gaseous and a
dark component to study the effects of a metagalactic UV background on
the formation of dwarf galaxies. In the case of an ionising background
consistent with the existing observational constraints no effect has
been found in halos with circular velocity above $60$~km~s$^{-1}$. The
effect is very mild in halos with $40 < V_{\mboxs{c}} <
60$~km~s$^{-1}$, and the reduction in the total gas mass which cools
below 8000~K never surpasses 30 \%. Below the circular velocity
40~km~s$^{-1}$ there is a sharp transition towards the case in which
the gas does not cool but it is gravitationally confined by the
potential well of the dark halo. Halos with a virial temperature below
the equilibrium photoionisation temperature cannot confine the gas.
\end{abstract}


\section{Introduction}

In the standard picture of structure formation smaller objects
collapse first, and merge later on to form bigger systems. As
the gas falls into the potential well of the forming halos it 
passes through a shock in which its bulk kinetic energy is 
converted into thermal energy. In Paper I \cite{fw} we studied 
how the hot primordial plasma can lose its thermal energy via 
radiative cooling. The gas cools more efficiently at higher 
redshifts, due to the bigger densities, and in smaller halos, due 
to line cooling. The result is then too many dwarf galaxies, which
causes the faint end of the luminosity function to be too steep
compared to observations (c.f.\ White \& Rees 1978; Kauffman, White \&
Guiderdoni 1993; Cole \itl{et al.} 1994).

Several possible solutions to the overcooling problem have been
contemplated in the literature. A first possibility is energy feedback
from the stars which form out of the cold dense gas sitting at the
centre of the halo. Dekel \& Silk (1986) argue that supernova
explosions could produce a big gas loss in halos up to a circular
velocity $V_{\mboxs{c}} \sim 100$~km~s$^{-1}$. However, the physics of
star formation and how stars interact with the surrounding
interstellar medium is still very poorly understood. Another
possibility is a higher merging rate, but then one could run into the
problem of thickening too much the discs of spiral galaxies (T\'{o}th
\& Ostriker 1992; Navarro et al.\ 1994; Vel\'{a}zquez \& White 1998). 
A third possibility was first suggested by Ikeuchi (1986) and Rees 
(1986) as a model for the Ly-$\alpha$ clouds: a UV background could 
prevent the primordial gas from cooling and sinking towards the centre
of small halos. Efstathiou (1992) pointed out that an ionising background
compatible with the existing observational constraints could reduce
the predicted number of dwarf galaxies.

These three possibilities are not exclusive, and the final answer to
the overcooling problem is probably a combination of all of them. In
the present paper I explore the role of the UV background in the
process of galaxy formation. The strongest piece of evidence for the
presence of a metagalactic ionising background is the absence of the 
so-called Gunn-Peterson trough \cite{gp} in the spectra of quasars 
(Steidel \& Sargent 1987; Schneider \itl{et al.} 1989; Bahcall \itl{et
al.} 1991; Webb \itl{et al.} 1992). Such a low Ly-$\alpha$ optical
depth could also be explained if there was no intergalactic medium 
(IGM), but this is quite improbable since the process of galaxy
formation is not thought to be 100 \% efficient. In addition, the
detection of intergalactic ionised helium absorption by the Hubble 
Space Telescope \cite{je} has not only confirmed the existence of a 
highly ionised IGM, but also the presence of a substantial amount of 
helium as predicted by the Big Bang nucleosynthesis.

The present paper is organised as follows: in Section 2 I briefly
describe the model assumed for the ionising spectrum; in Section 3 I
present the numerical simulations done with the spherical Lagrangian 
code described in Forcada-Mir\'{o} \& White (1998); the results obtained
are discussed in Section 4. In all cases the Hubble constant has been 
taken to be $H_{o} = 50$~km~s$^{-1}$~Mpc$^{-1}$, and the baryonic
content of the universe is ${\Omega}_{\mboxs{B}} = 0.05$.

\section{Model for the UV background}

In the simulations presented in this paper I  have assumed an
ionising spectrum which is a simple power law,
\begin{equation}
J_{\nu} = J_{{\nu}_{\mboxs{L}}}(z) {\left( \frac{\nu}{{\nu}_{\mboxs{L}}}
\right)}^{- \alpha} \; ,
\end{equation}
where $\alpha$ is the spectral index, and $J_{{\nu}_{\mboxs{L}}}$
is the mean intensity of the radiation field at the Lyman limit. The
ionising background is switched on at a redshift $z_{\mboxs{ion}}$,
and its subsequent dependence on the redshift is taken as in 
Efstathiou (1992),
\begin{equation}
J_{{\nu}_{\mboxs{L}}}(z) = \left\{ \begin{array}{ll} 0 & z >
z_{\mboxs{ion}} \\ \frac{J_{-21}}{1 + {[5/(1+z)]}^{4}} & z <
z_{\mboxs{ion}} \end{array} \right. \; ,
\end{equation}
where $J_{-21}$ is in units of $10^{-21}$ ergs~cm$^{-2}$~s$^{-1}$~Hz
$^{-1}$~sr$^{-1}$.

The proximity effect provides an estimate of the intensity of the UV
background at the Lyman limit \cite{lw},
\begin{eqnarray}
\log J_{{\nu}_{\mboxs{L}}} = -21 \pm 0.5 & \mbox{for} \; 
1.7 < z < 3.8 \; ,
\end{eqnarray}
where $J_{{\nu}_{\mboxs{L}}}$ is in units of
ergs~cm$^{-2}$~s$^{-1}$~Hz$^{-1}$~sr$^{-1}$. More recent
determinations based on high resolution spectra do not find also
significant evidence for a change in $J_{{\nu}_{\mboxs{L}}}$ in the
redshift range $2 < z < 4$, but they seem to favour a smaller value
$\sim 3 \times 10^{-22}$~ergs~cm$^{-2}$~s$^{-1}$~Hz$^{-1}$~sr$^{-1}$
(Willinger \itl{et al.} 1994; Fern\'{a}ndez-Soto \itl{et al.} 1995;
Giallongo \itl{et al.} 1996; Cooke \itl{et al.} 1997). At lower 
redshifts $J_{{\nu}_{\mboxs{L}}}$ is still poorly constrained. Kulkarni
\& Fall (1993) have analysed HST observations, and obtain a value
$\sim 6 \times 10^{-24}$~ergs~cm$^{-2}$~s$^{-1}$~Hz$^{-1}$~sr$^{-1}$. 
All these estimates are determined assuming a power-law distribution 
in redshift of the Ly-$\alpha$ clouds, and it could be the case that 
such a distribution is not adequate. In the simulations I have
considered a stronger background, with $J_{-21} = 1$, and a weaker
background, with $J_{-21} = 0.1$. The existing observational
constraints fall in between these two cases.

A possible source for the ionising radiation is quasars, which have a
spectrum that in the extreme ultraviolet is well approximated by a
simple power law $\alpha \approx 1.8$ \cite{zk}. However, the spectrum
of quasars is too hard to be in agreement with HeII observations
(Madau \& Meiksin 1994; Savaglio \itl{et al.} 1997). Also if the
population of quasars has a strong decline at $z > 3$ the emitted UV
radiation is insufficient to ionise the medium to the required level
\cite{mo1}. Recent redshift surveys seem to indicate that the decline
in the quasar population is not so strong (Boyle 1991; Irwin \itl{et
al.} 1991), and then, the ionising flux provided by the quasars might
be sufficient \cite{mo2}. Another possible source of UV radiation is 
young stars in galaxies, which only emit below the HeII ionisation
edge. Therefore models including the contribution from both quasars 
and starbursting galaxies provide a softer spectrum than models with
quasars only \cite{mo1}. Also the first generation of baryonic objects
in which stars could form might emerge at redshifts $z \approx 10-50$ 
thanks to the formation of molecular hydrogen \cite{hl}. Therefore, 
I have considered both a hard background, with $\alpha = 1.5$, and a 
soft background, with $\alpha = 5$, and the ionising flux is switched
on at two different epochs, $z_{\mboxs{ion}} = 5$ and $z_{\mboxs{ion}}
= 10$.

In the numerical simulations presented in the following section we
refer as a fiducial model to the case with a UV background which has
$J_{-21} = 1$, $\alpha = 1.5$, and $z_{\mboxs{ion}} = 5$.

\section{Numerical results}

\subsection{Effects on halos with circular velocity $V_{\mboxs{c}} > 
40$~km s$^{-1}$}

In the presence of the fiducial UV background the behaviour of the 
radiative shocks differs significantly from the behaviour discussed 
in Paper I up to a circular velocity $V_{\mboxs{c}} \approx 
60$~km~s$^{-1}$. Shortly after the ionizing background is switched 
on at $z_{\mboxs{ion}} = 5$ there is a short period of time in which 
the shock wave moves outwards very fast, approaching the nonradiative 
behaviour (Figure 1). This effect is stronger as one goes towards 
lower circular velocities. Notice also that the oscillations with 
growing period and amplitude which appeared in the simulations with 
no ionizing background are not present in these calculations.

In Figure 2 I compare the effect of varying some of the parameters
of the model in the case of a halo with circular velocity
$V_{\mboxs{c}} = 50$~km~s$^{-1}$. If the background is switched on at 
a higher redshift (i.e.\ $z_{\mboxs{ion}} = 10$) the shock approaches 
the non-radiative behaviour at an earlier time, but the subsequent
evolution is very similar to the fiducial model. If the intensity of
the ionizing radiation is decreased (i.e.\ $J_{-21} = 0.1$) some small
oscillations appear in the evolution of the shock wave. The most
drastic effect is obtained by softening the spectrum (i.e.\ $\alpha =
5$), since the oscillations with growing period and amplitude are 
present again.

On the other hand, the evolution of the total mass which has been 
able to cool below 8000~K does not differ so much from the evolution 
when $J_{-21} = 0$. For redshifts $3 < z < 5$ the presence of the UV
background does not affect at all the ability of the gas to cool 
inmediately after being shock-heated. It is not until redshifts 
$z < 3$ than the mass of cold gas is smaller in the presence of the 
fiducial UV background, but in any case, the reduction does not 
surpass 30 \% (Figure 3). Ionizing the gas at a higher redshift or 
reducing the intensity of the background does not affect practically 
the amount of cold gas. However, in the case of a softer spectrum, the
evolution of the cold mass is closer to the evolution in the absence
of the ionizing background.

\subsection{Effects on halos with circular velocity $V_{\mboxs{c}} < 
40$~km s$^{-1}$}

I have simulated as well a halo with circular velocity $V_{\mboxs{c}} 
= 30$~km~s$^{-1}$. In this case the evolution of the system is more 
complicated and it is better understood by looking at the radial 
profiles of the velocity, density and temperature at different 
redshifts. In Figure 4 one can see that an outflow forms, 
which is driven by the photoionizing radiation. As the system 
evolves the photoionization-driven wind gradually dies out 
and the cloud starts growing via cosmological accretion again. 
The result is then a cloud in quasi-hydrostatic equilibrium in 
which hot gas is confined by the gravity of the dark minihalo. 
Shortly after switching on the ionising background the gas cannot 
cool and settle towards the centre of the system anymore.

As in the case for bigger halos, these results are not very 
sensitive to reionizing the gas at an earlier time. If the flux 
of the UV background is reduced the wind is much weaker and the 
gas has some residual infall. In the case of a much softer spectrum 
the ionizing radiation cannot drive a wind, and the gas is able to 
fall in the potential well of the halo and cool efficiently.

\section{Discussion}

There are two different ways in which the UV background can influence
the ability of the primordial plasma to cool and collapse.  One of
them is by increasing the ionised fraction of the gas, and thus,
reducing line cooling. The other is by heating the IGM by
producing electrons with excess kinetic energy.

The efficiency of the ionising radiation to modify the emissivity of 
the plasma depends on the hardness of the spectrum and the density of 
the plasma (Figure 5). If the spectrum is hard both hydrogen and 
helium are strongly ionised, and the cooling function does not have 
such a negative power-law dependence on the temperature, but a soft 
spectrum does not affect the ionisation state of helium. This explains
why in the case of a hard spectrum the shock wave is not oscillatory
unstable, but the oscillations are present if the spectrum is soft.
On the other hand, collisional ionisation rates are proportional 
to ${{\rho}_{g}}^{2}$, since they are two-body processes, while 
photoionisation rates are only proportional to ${\rho}_{g}$. In this
way, if the density is low enough, line cooling is heavily suppressed,
and the emissivity of the plasma is reduced up to temperatures 
$T \sim 10^{6}$~K. However, at the post-shock density, which is on 
average of the order of 120 times the background density \cite{fw}, 
the emissivity is only reduced up to a temperature few times
$10^{5}$~K. This explains why no significant effect is obtained for 
halos with a circular velocity $V_{\mboxs{c}} > 60$~km~s$^{-1}$. 
The gas is also able to cool efficiently enough in halos with
circular velocity down to $V_{\mboxs{c}} \approx 40$~km~s$^{-1}$, 
and thus, there is not so much difference in the total cold mass. 
However, the reduction in the emissivity of the plasma increases 
the cooling length and the shock wave moves outwards faster until 
the free-fall time is of the order of the cooling time (see Paper I).

There is a temperature $T_{\mboxs{eq}}$, known as the photoionisation 
equilibrium temperature, at which the energy losses $\Lambda$ equal 
the heating due to the photoionising radiation $\Gamma$. As one goes 
towards smaller halos the virial temperature is only slightly bigger 
than the equilibrium temperature, which in the fiducial case is 
$T_{\mboxs{eq}} \approx 3 \times 10^{4}$~K. Consequently, in halos 
with a circular velocity in the range 25-35 km~s$^{-1}$ the gas is 
still gravitationally confined, but the heating due to the
photoionising keeps it at a temperature of the order of 
$T_{\mboxs{eq}}$. Such systems were proposed by Ikeuchi (1986) and 
Rees (1986) as possible candidates for the Ly-$\alpha$ forest. This 
velocity range is in agreement with Meiksin (1994), and is very much 
independent of $z_{\mboxs{ion}}$ and $J_{\mboxs{-21}}$. However, in 
the case of a softer spectrum the equilibrium temperature is only 
$10^{4}$~K, and the Ly-$\alpha$ clouds would have a characteristic 
circular velocity $V_{\mboxs{c}} \sim 20$~km~s$^{-1}$.

Finally, I would like to point out that a time-dependent calculation 
of the ionisation state of the plasma is crucial, since calculations 
assuming equilibrium abundances underestimate the temperature of the 
intergalactic medium by nearly an order of magnitude in the early 
stages of reionisation. This is because the photoionisation time is 
much shorter than the recombination time just after the background 
is switched on, and the ionising radiation can deposit energy more 
efficiently in the plasma. At later times, radiative recombination 
becomes important, and both calculations converge.

\section{Conclusions}

The qualitative conclusion I draw out of the simulations is that the
metagalactic UV background has a very mild effect on the formation of
dwarf galaxies. The way the ionising radiation affects the collapse of
small objects is more by heating the pre-shock plasma than by
modifying the cooling function. Our results are only marginally in
agreement with the spherical collapse calculations of Thoul \&
Weinberg (1996), which indicated a reduction $\sim 50 \%$ of the 
cooled mass fraction in halos with $V_{\mboxs{c}} \sim
50$~km~s$^{-1}$, and a negligible effect above $V_{\mboxs{c}} \sim 
75$~km~s$^{-1}$. They find a stronger effect probably because they 
use a slightly harder spectrum (i.e.\ $\alpha = 1$), and as I have 
showed the effect of the UV background is very sensitive to the
spectral index. On the other hand, the three-dimensional simulations 
of Navarro \& Steinmetz (1996) show a mild effect of the metagalactic 
background all the way up to normal galaxies (i.e.\ $V_{\mboxs{c}} 
\sim 200$~km~s$^{-1}$). As the authors themselves suggest, this must 
be due to the fact that a fraction of the gas which falls onto the 
potential well of a bigger halo collapsed at an earlier time onto 
smaller halos.

In any case, all the theoretical work seems to confirm that a UV 
metagalactic background is far from being enough to solve the 
overcooling problem, since the luminosity function semi-analytic 
models predict differs from observations at circular velocities 
$V_{\mboxs{c}} < 100$~km~s$^{-1}$.

\section*{ACKNOWLEDGMENT}

I would like to thank the Max-Planck Gesellschaft for the economical 
support. I am also grateful to Tom Abel, Martin Haehnelt, Martin Rees
and Simon White for many helpful discussions.

\clearpage 

\begin{figure*}
   \cidfig{17cm}{20}{150}{570}{700}{\DIRFIGS shockp1.ps}
\vspace{6pt}{
{\bf Figure 1:} Evolution in time of the shock wave position 
(solid line) and the cooling wave position (crosses) for halos 
with $40 < V_{\mboxs{c}} < 60$~km~s$^{-1}$ in the presence of 
the fiducial UV background. Long-dashed lines correspond to the 
adiabatic evolution of the shock.
}
\end{figure*}

\clearpage

\begin{figure*}
   \cidfig{17cm}{20}{150}{570}{700}{\DIRFIGS shockp2.ps}
\vspace{6pt}{
{\bf Figure 2:} Evolution in time of the shock wave position 
(solid line) and the cooling wave position (crosses) for a halo 
with $V_{\mboxs{c}} = 50$~km~s$^{-1}$ in the presence of 
different UV backgrounds. Long-dashed lines correspond to the
adiabatic evolution of the shock.
}
\end{figure*}

\clearpage

\begin{figure*}
   \cidfig{17cm}{20}{150}{570}{700}{\DIRFIGS massp1.ps}
\vspace{6pt}{
{\bf Figure 3:} Mass which has been able to cool below 8000~K for 
halos with $V_{\mboxs{c}} = 40, \; 50, \; 60$ km~s$^{-1}$. Solid line:
Evolution in the absence of a UV background. Dashed line: Evolution in the 
presence of the fiducial UV background. Dotted-line: Reduction of a 30
\%.
}
\end{figure*}

\clearpage

\begin{figure*}
   \cidfig{17cm}{20}{150}{570}{700}{\DIRFIGS profp1.ps}
\vspace{6pt}{
{\bf Figure 4:} Radial profiles at different redshifts 
for a halo with $V_{\mboxs{c}} = 30$~km~s$^{-1}$ in the presence of
the fiducial UV. Top: Velocity. Middle: Density. Bottom: Temperature. 
The variables have been scaled as in Paper I.
}
\end{figure*}

\clearpage

\begin{figure*}
   \cidfig{8.5cm}{20}{425}{575}{700}{\DIRFIGS heat.ps}
   \vspace{8.5cm}
\vspace{6pt}{
{\bf Figure 5:} Cooling and heating functions for a primordial
gas. In all cases $Y_{\mboxs{P}} = 0.24$ and $J_{-21} = 1$. Compton
cooling is not considered. Left panel: Effect of density. Thin lines 
correspond to ${\rho}_{g} = 5.864 \times 10^{-29}$~gr~cm$^{-3}$. Thick
lines correspond to a density 200 times bigger. In both cases the spectral 
index $\alpha = 1.5$. Right panel: Effect of spectral index. Thin
lines correspond to $\alpha = 1.5$. Thick lines correspond to $\alpha
= 5$. In both cases ${\rho}_{g} = 5.864 \times 10^{-29}$~gr~cm$^{-3}$.
}
\end{figure*}


\begin{thebibliography}{}
 \bibitem[\protect\citename{Boyle }1991]{bo} Boyle, B.J., 1991,
 Cosmology and Fundamental Physics, Texas/ESO-CERN Symposium on
 Relativistic Astrophysics, ed.\ J.D. Barrow, L. Mestel, P.A. Thomas
 \bibitem[\protect\citename{Cole \itl{et al.} }1994]{ca} Cole S.,
 Arag\'{o}n-Salamanca A., Frenk C.S., Navarro J.F., Zepf S.E., 1994,
 MNRAS, 271, 781
 \bibitem[\protect\citename{Cooke \itl{et al.} }1997]{ce} Cooke, A.J.,
 Espey, B., Carswell, R.F., 1997, MNRAS, 284, 552
 \bibitem[\protect\citename{Dekel \& Silk }1986]{ds} Dekel A., Silk
 J., 1986, ApJ, 303, 39
 \bibitem[\protect\citename{Efstathiou }1992]{ef} Efstathiou, G.,
 1992, MNRAS, 256, 43P
 \bibitem[\protect\citename{Fern\'{a}ndez-Soto \itl{et al.} }1995]{fb}
 Fern\'{a}ndez-Soto, A., Barcons, X., Carballo, R., Webb, J.K., 1995,
 MNRAS, 277, 235
 \bibitem[\protect\citename{Forcada-Mir\'{o} \& White }1998]{fw}
 Forcada-Mir\'{o}, M.I., White, S.D.M., 1998, submitted to MNRAS
 (Paper I)
 \bibitem[\protect\citename{Giallongo \itl{et al.} }1996]{gc}
 Giallongo, E., Cristiani, S., D'Odorico, S., Fontana, A., Savaglio,
 W., 1996, ApJ, 446, 46
 \bibitem[\protect\citename{Gunn \& Peterson }1965]{gp} Gunn, J.E.,
 Peterson, B.A., 1965, ApJ, 142, 1633
 \bibitem[\protect\citename{Haiman \& Loeb }1997]{hl} Haiman, Z.,
 Loeb, A., 1997, ApJ, 483, 21
 \bibitem[\protect\citename{Ikeuchi }1986]{ik} Ikeuchi, S., 1986,
 ApSS, 118, 509
 \bibitem[\protect\citename{Irwin \itl{et al.} }1991]{im} Irwin, M.,
 McMahon, R.G., Hazard, C., 1991, The Space Distribution of Quasars,
 ASP Conf.\ Ser.\ 21 (San Francisco: ASP)
 \bibitem[\protect\citename{Jackobsen \itl{et al.} }1994]{je}
 Jackobsen, P. \itl{et al.}, 1994, Nature, 370, 35
 \bibitem[\protect\citename{Kauffmann  \itl{et al.} }1993]{kw}
 Kauffmann G., White S.D.M., Guiderdoni B., 1993, MNRAS, 264, 201
 \bibitem[\protect\citename{Kulkarni \& Fall }1993]{kf} Kullkarni, P.,
 Fall, S.M., 1993, ApJ, 413, L67
 \bibitem[\protect\citename{Lu \itl{et al.} }1991]{lw} Lu, L., Wolfe,
 A.M., Turnshek, D.A., 1991, ApJ, 367, 19
 \bibitem[\protect\citename{Madau \& Meiksin }1994]{mm1} Madau, P.,
 Meiksin, A., 1994, ApJ, L53
 \bibitem[\protect\citename{Meiksin }1994]{me} Meiksin, A., 1994, ApJ,
 431, 109
 \bibitem[\protect\citename{Miralda-Escud\'{e} \& Ostriker }1990]{mo1}
 Miralda-Escud\'{e}, J., Ostriker, J.P., 1990, ApJ, 350, 1
 \bibitem[\protect\citename{Miralda-Escud\'{e} \& Ostriker }1992]{mo2}
 Miralda-Escud\'{e}, J., Ostriker, J.P., 1992, ApJ, 392, 23
 \bibitem[\protect\citename{Navarro \itl{et al.} }1994]{nf} Navarro,
 J.F., Frenk, C.S., White, S.D.M., 1994, MNRAS, 267, L1
 \bibitem[\protect\citename{Navarro \& Steinmetz }1997]{ns} Navarro,
 J.F., Steinmetz, M., 1997, ApJ, 478, 13
 \bibitem[\protect\citename{Rees }1986]{re} Rees, M.J., 1986, MNRAS,
 218, 25P
 \bibitem[\protect\citename{Savaglio \itl{et al.} }1997]{sc} Savaglio,
 S., Cristiani, S., D'Odorico, S., Fontana, A., Giallongo, E., Molaro,
 P., 1997, A\&A, 318, 347
 \bibitem[\protect\citename{Schneider \itl{et al.} }1989]{sg}
 Schneider, D.P., Schmidt, M., Gunn, J.E., 1989, AJ, 98, 1951
 \bibitem[\protect\citename{Steidel \& Sargent }1987]{ss} Steidel,
 C.C., Sargent, W.L.W.
 \bibitem[\protect\citename{Thoul \& Weinberg }1996]{tw} Thoul, A.A.,
 Weinberg, D.H., 1996, ApJ, 465, 608
 \bibitem[\protect\citename{T\'{o}th \& Ostriker }1992]{to} T\'{o}th,
 G., Ostriker, J.P., 1992, ApJ, 389, 5
 \bibitem[\protect\citename{Vel\'{a}zquez \& White }1997]{vw} 
 Vel\'{a}zquez, H., White, S.D.M., 1998, MNRAS, in press
 \bibitem[\protect\citename{Webb \itl{et al.} }1992]{wb} Webb, J.K.,
 Barcons, X., Carswell, R.F., Parnell, H.C., 1992, MNRAS, 255, 319
 \bibitem[\protect\citename{White \& Rees }1978]{wr} White, S.D.M.,
 Rees, M.J., 1978, MNRAS, 183, 341
 \bibitem[\protect\citename{Willinger \itl{et al.} }1994]{we}
 Willinger, G.M. \itl{et al.}, 1994, ApJ, 428, 574
 \bibitem[\protect\citename{Zheng \itl{et al.} }1997]{zk} Zheng, W., 
 Kriss, G.A., Telfer, R.C., Grimes, J.P., Davidsen, A.F., 1997, ApJ,
 475, 469
\end{thebibliography}
\end{document}